\documentclass[12pt]{article}
\begin{document}
\title {One particle quantum equation in a de Sitter spacetime} 
\date{} 
\author{R. A. Frick\thanks{Email: rf@thp.uni-koeln.de}\\ Institut f\"ur Theoretische Physik, Universit\"at K\"oln,\\ Z\"ulpicher Str. 77, 50937 K\"oln, Germany}

\maketitle
\begin{abstract}
We consider a free particle in a de Sitter spacetime. We use a picture in which the analogs of the Schr\"odinger operators  of the particle are independent of  both the time and the space coordinates. These operators  induce  operators which are related to Killing vectors of the  de Sitter spacetime.  
\end{abstract}

\section{Introduction}

In this paper we  use a generalized Schr\"odinger  (GS) picture to describe  a  particle in a de Sitter spacetime.  The GS picture is based on the principal series of the infinite-dimensional unitary irreducible representations (UIR) of the Lorentz group \cite{Joos} and a spacetime transformation. The principal series  correspond to the eigenvalues $1+\alpha^2-{\lambda}^2,\,(0\leq\alpha<\infty,\quad\lambda=-s,...,s,{\quad}s = spin)$) of the first ${C_1}={\bf N}^2-{\bf J}^2$, (${\bf N},\,{\bf J}$ are boost and rotation generators) and the eigenvalues $\alpha\lambda$ of the second Casimir operator ${C_2}={\bf N}\cdot{\bf J}$ of the Lorentz group. The representations $(\alpha,\lambda)$ and $(-\alpha,-\lambda)$ are unitarily equivalent. In the momentum representation (${\bf p}$ = momentum, $p_0=\sqrt{m^2c^4+c^2{\bf p}^2}$, $ m=mass$, $s=0$, ${\bf n}^{2}({\theta},{\varphi})=1$),  the  operator ${C_1}({\bf p})$ has the eigenfunctions 
\begin{equation}
\label{1.1}
\xi({\bf p},{\alpha},{\bf n}):=[(p_{0}-c{\bf p}\cdot{\bf n})/mc^2]^{-1+i\alpha},\end{equation}

 In \cite{Shap}, it was proposed to classify the states of a relativistic particle by means of  the operators ${C_1}$ and ${C_2}$,  and to carry out an expansion  known as Shapiro transformations or the expansion of the Lorentz group 
\begin{equation}
\label{1.2}
\psi({\alpha},{\bf n})=\frac{1}{(2\pi)^{3/2}}
\int\frac{d{\bf p}}{p_0}\,\psi({\bf p})\,\xi^{\ast}({\bf p},{\alpha},{\bf n}).
\end{equation}
Here $\psi({\bf p})$ denotes the wave function of a spinless  particle in the momentum space representation and $\psi({\alpha},{\bf n})$ the wave function of the particle in the ${\alpha}{\bf n}$ representation. The expansion (\ref{1.2})  does not include any dependence on the time  and the space coordinates $t$,  ${\bf x}$, i.e. it is ``spacetime independent''.  In \cite{Kad} the expansion over the functions $\xi^{\ast}({\bf p},{\alpha},{\bf n})$ was used to introduce the ``relativistic configurational'' representation (in following $\rho{\bf n}$-representation, $\rho=\alpha\hbar/mc$) in the framework of a two-paricle equation of the quasipotential type. In this approach the variable $\rho$ was interpeted as the relativistic generalization of a relative coordinate. In \cite{Kad} it was shown that the corresponding operators of the Hamiltonian $H(\rho,{\bf n})$ and the 3-momentum ${\bf P}(\rho,{\bf n})$, defined on the functions $\xi^{\ast}({\bf p},{\rho},{\bf n})$, has a form of the differential-difference operators (their explicit form shall be used in what follows, see Eqs.(\ref{1.32}) and (\ref{1.33})). Various applications of the unitary representations of the Lorentz group and the $\rho{\bf n}$-representation can be found in the literature.

In our previous papers \cite{Fri1} it has been shown that the UIR of the Lorentz group and the expansion (\ref{1.2}) may be also used in a so-called generalized Schr\"odinger picture in which the analogs of the Schr\"odinger operators of a particle are independent of both the time and the space coordinates. It was found that the operators $H(\rho,{\bf n})$,  ${\bf P}(\rho,{\bf n})$,  ${\bf J}({\bf n})$ and ${\bf N}(\rho,{\bf n})={\rho}{\bf n}+({\bf n}\times{\bf J}-{\bf J}\times{\bf n})/2mc$ satisfy the commutations relations of the Poincar\'e algebra in the $\rho{\bf n}$-representation. In the GS picture we have two spacetime independent representations of the Poincar\'e algebra; the ${\bf p}$ and the $\rho{\bf n}$-representation.  This leads to  another Fourier transformation to that in the  standard quantum theory. For a free particle in the Minkowski spacetime  the coordinates $t$, ${\bf x}$ may be introduced in the states  with the help of the transformation  
\begin{equation}
\label{1.3}
 S(t,{\bf x})=\exp[-i(tH-{\bf x}\cdot{\bf P})/\hbar],
\end{equation}   
where $H$ and ${\bf P}$ are the Hamilton and momentum operators of the particle in the $\rho{\bf n}$ or in the ${\bf p}$-representation. Instead of (\ref{1.2}) we obtain
\begin{equation}
\label{1.4}
S(t,{\bf x})\psi({\rho},{\bf n})=\psi({\rho},{\bf n},t,{\bf x})=\frac{1}{(2\pi)^{3/2}}
\int\frac{d{\bf p}}{p_0}\,\psi({\bf p},t,{\bf x})\,\xi^{\ast(0)}({\bf p},{\rho},{\bf n}),
\end{equation}
where $\psi({\bf p},t,{\bf x})=\psi({\bf p})\exp[-ix{\cdot}p/\hbar]$. It follows that in the GS picture the  spacetime coordinates appear in the states in  the $\rho{\bf n}$ and in the ${\bf p}$-representation. The use of the $\rho{\bf n}$-representation in this picture  makes it possible to describe extendent objects like strings.

For the functions $\psi({\rho},{\bf n},t,{\bf x})$ and $\psi({\bf p},t,{\bf x})$ we have
\begin{equation}
\label{1.5}
i\hbar\frac{\partial}{\partial{t}}\psi=H\psi,\quad{-i}\hbar\frac{\partial}{\partial{\bf x}}\psi={\bf P}\psi.
\end{equation} 
In this  equations the operators $H$ and ${\bf P}$ induce on the left-hand-side the operators of the Lie-algebra of Killing vectors of the  Minkowski spacetime. A full Poincar\'e algebra contain in addition to the operators in (\ref{1.5}) the Lorentz rotation and boost generators.  For the first Casimir operator of the Poincar\'e group  we have
\begin{equation}
\label{1.6}
{C}(t,{\bf x})\psi({\bf p},t,{\bf x})={C}({\bf p})\psi({\bf p},t,{\bf x}),{\quad}{C}(t,{\bf x})\psi({\rho},{\bf n},t,{\bf x})={C}(\rho{\bf n})\psi({\rho},{\bf n},t,{\bf x}).
\end{equation}
In the paper \cite{Fri2}, it was shown that equations like (\ref{1.5})  may be used to describe a particle in an Anti-de Sitter spacetime. It was found that operators with an external field in the ${\rho}{\bf n}$-representation  which corresponds to an attractive force induce operators ${K}_a(t,x^{i})$ which are related to Killing vectors  of the AdS spacetime $(a=1,2,...10;$ $\lbrace{x^{i}\rbrace}$, $i=1,2,3.)$
\begin{equation}
\label{1.7}
{K}_a(t,x^{i})\widetilde{\Phi}({\rho},{\bf n},t,x^{i})=B _a({\rho},{\bf n})\widetilde{\Phi}({\rho},{\bf n},t,x^{i}).
\end{equation}
Here $\widetilde{\Phi}$ denotes the wave function of the particle in  the $\rho{\bf n}$-representation. The operators  ${K}_a(t,x^{i})$ satisfy  the same commutation rules as  the spacetime independent operators $B _a({\rho},{\bf n})$ , except for the minus signs on the right-hand sides. In the framework of GS-picture, a particle in a spacetime is free if a set of operators in the $\rho{\bf n}$ or in the ${\bf p}$-representation force the introduction of  Killing vector fields of this spacetime. 

In the present paper we will show that  an external massless field in the  $\rho{\bf n}$ or in the momentum representation which corresponds to a repulsive force induce  operators which are related to Killing vectors of the de Sitter spacetime. We want to show that a particle in a de Sitter  spacetime may be described by the equations like (\ref{1.7}). The external fields in the ${\bf p}$ or in the  $\rho{\bf n}$-representation do violate the commutation relations of the Poincar\'e algebra. We have the problem of determining  observables in the GS picture. Here we  use only the coordinate system which corresponds to an exponentially expanding world \cite{Bir,Stro,Obu}. We shall show that in this coordinates  the spacetime independent momentum operators of the particle in addittion to the external field contain the Lorentz boost generators.  In Sect.2 we study the  motion of a particle in a two dimensional de Sitter spacetime. We use the ${\rho}$-representation. In Sect.3 the four dimensional de Sitter spacetime and the $\rho{\bf n}$-representation  are used. In Appendix we use the momentum representation.
\section{Motion in a two dimensional dS spacetime}

In a  two dimensional Minkowski spacetime the motion  of a free particle is described by the equations
\begin{equation}
\label{1.8}
i\hbar\frac{\partial}{\partial{t}}\psi(\rho,t,x)=H(\rho)\psi(\rho,t,x),\quad{-i}\hbar\frac{\partial}{\partial{x}}\psi(\rho,t,x)=P(\rho)\psi(\rho,t,x),
\end{equation}
where the Hamilton operator $H(\rho)$ and the momentum operator $P(\rho)$ have the form
\begin{equation}
\label{1.9}
H({\rho})=mc^2\cosh(-\frac{i\hbar}{mc}{\partial}_{{\rho}}),{\quad}P({\rho})=mc\sinh(-\frac{i\hbar}{mc}{\partial}_{{\rho}}).
\end{equation}
 The operators $H(\rho)$, $P(\rho)$  and the operator $N(\rho)$  ($N(\rho)=\rho)$ satisfy the commutation relations of the Poincar\'e algebra,
\begin{equation}
\label{1.10}
\lbrack{N},P\rbrack=i\frac{\hbar}{mc^2}H,\quad\lbrack{P},{H}\rbrack=0,\quad\lbrack{H},{N}\rbrack=-i\frac{\hbar}{m}P.
\end{equation}
In (\ref{1.6}), in a model of a relativistic oscillator we have used an external field which corresponds to an attractive force $(\omega$ = frequency)
\begin{equation}
\label{1.11}
\widetilde{H}^{'}(\rho)=\frac{m{\omega}^2}{2}\rho\left(\rho-i\frac{\hbar}{mc}\right)e^{-i\frac{\hbar}{mc}{\partial}_{\rho}},{\quad}\widetilde{P}_1^{'}(\rho)=\frac{m{\omega}^2}{2c}\rho\left(\rho-i\frac{\hbar}{mc}\right)e^{-i\frac{\hbar}{mc}{\partial}_{\rho}}.
\end{equation}
 The operators  
\begin{equation}
\label{1.12}
\hat{P_0}(\rho)=H({\rho})+\widetilde{H}^{'}(\rho),{\quad}\hat{P_1}(\rho)=P({\rho})+\widetilde{P}_1^{'}(\rho),
\end{equation}
and $\rho$ satisfy the commutation relations 
\begin{equation}
\label{1.13}
\lbrack\rho,\hat{P_1}\rbrack=i\frac{\hbar}{mc^2}\hat{P_0},\quad\lbrack\hat{P_1},\hat{P_0}\rbrack=-i\hbar{m}{\omega}^2\rho,\quad\lbrack\hat{P_0},\rho\rbrack=-i\frac{\hbar}{m}\hat{P_1}.
\end{equation}

In the nonrelativistic limit the operators $\hat{P_0}(\rho)-mc^2$ and $\hat{P_1}(\rho)$ assume the form
\begin{equation}
\label{1.14}
\hat{P_0}_{\rm{nr}}=-\frac{{\hbar}^2}{2m}\frac{{\partial}^2}{\partial{\rho}^2}+\frac{m{\omega}^2}{2}{\rho}^2,{\quad}\hat{P_1}_{\rm{nr}}=-i\hbar\frac{\partial}{\partial\rho}.
\end{equation} 
It was found that the operators  $\hat{P_0}$,  $\hat{P_1}(\rho)$ and $\rho$ induce  operators which are related to  the Killing vectors of an AdS spacetime.

Here, in analogy to (\ref{1.11}) for the external field which corresponds to a repulsive force  we use the following operators
\begin{equation}
\label{1.15}
H^{'}_0(\rho)=-\frac{mc^2}{2{\ell}^2}\rho\left(\rho-i\frac{\hbar}{mc}\right)e^{-i\frac{\hbar}{mc}{\partial}_{\rho}},{\quad}P^{'}_1(\rho)=-\frac{mc}{2{\ell}^2}\rho\left(\rho-i\frac{\hbar}{mc}\right)e^{-i\frac{\hbar}{mc}{\partial}_{\rho}}
\end{equation}
where the constant $\ell$  has the dimension of length. Here the quantity $\ell$ is related to the radius of a two dimensional de Sitter spacetime.

The operators
\begin{equation}
\label{1.16}
{\Pi_0}(\rho)=H(\rho)+H^{'}_0(\rho),\quad{\Pi_1}(\rho)=P(\rho)+P^{'}_1(\rho),
\end{equation}
satisfy the commutation relations 
\begin{equation}
\label{1.17}
{\lbrack}{\rho},{\Pi_1}\rbrack=i\frac{\hbar}{mc^2}{\Pi_0},\quad\lbrack{\Pi_1},{\Pi_0}\rbrack=i\hbar{m}{\frac{c^2}{\ell^2}}{\rho},\quad\lbrack{\Pi_0},{\rho}\rbrack=-i\frac{\hbar}{m}{\Pi_1}.
\end{equation}
The Casimir operator is a multiple of the identity operator ${I}$
\begin{equation}
\label{1.18}
C(\rho)=-\frac{{\ell}^2}{(\hbar{c})^2}{\lbrace}{\Pi_0}^{2}-c^2{\Pi}_1^{2}{\rbrace}-\frac{m^2c^2}{{\hbar}^2}{\rho}^2=-\left(\frac{{m^2c^2\ell}^2}{{\hbar}^2}\right){I}.
\end{equation}

 Now in accordance with (\ref{1.7}),  we  introduce three operators ${\cal}K_{10}$, $K_{12}$ and  $K_{02}$  in terms of the spacetime coordinates of a two dimensional de Sitter spacetime with  the  same commutation rules  as  the operators ${\Pi_0}$, ${\Pi_1}$ and $\rho$,  except for the minus signs on the right-hand sides  ($d=2$)
\begin{equation}
\label{1.19}
\lbrack{K_{10},K_{1d}}\rbrack=-\frac{\imath\hbar}{mc^2}K_{0d},{\quad}\lbrack{K_{1d},K_{0d}}\rbrack=-\imath\hbar{m}\frac{c^2}{{\ell}^2}{K_{10}},{\quad}\lbrack{K_{0d},K_{10}}\rbrack=\frac{\imath\hbar}{m}{K_{1d}}.
\end{equation}
For the operators $K_{0d}$, $K_{1d}$ and $K_{10}$ we choose the  following realisation
\begin{equation} 
\label{1.20}
K_{0d}(t,x)=i{\hbar}(\frac{\partial}{\partial{t}}-\frac{cx}{{\ell}}\frac{\partial}{\partial{x}}),
\end{equation}
\begin{equation} 
\label{1.21}
K_{1d}(t,x)=i\frac{\hbar}{\ell}[x\frac{\partial}{\partial{ct}}-(\frac{{x^2}}{2\ell}+\frac{\ell}{2}e^{-2ct/\ell})\frac{\partial}{\partial{x}}-\frac{\ell}{2}\frac{\partial}{\partial{x}}],
\end{equation}
\begin{equation} 
\label{1.22}
K_{01}(t,x)=i\frac{\hbar}{mc}[-x\frac{\partial}{\partial{ct}}+(\frac{{x^2}}{2\ell}+\frac{\ell}{2}e^{-2ct/\ell})\frac{\partial}{\partial{x}}-\frac{\ell}{2}\frac{\partial}{\partial{x}}].
\end{equation}
These operators are related to Killing vectors of the de Sitter spacetime with metric 
\begin{equation}
\label{1.23}
ds^2=c^2dt^2-e^{2ct/{\ell}}dx^2.
\end{equation}
Using the relations (\ref{1.17}) and (\ref{1.19}) we find that a free particle in dS spacetime (\ref{1.23}) is describe by the quantum equations
\begin{equation}
\label{1.24}
K_{0d}(t,x)\Phi(\rho,t,x)={\Pi_0}\Phi(\rho,t,x),{\quad}K_{1d}(t,x)\Phi(\rho,t,x)={\Pi_1}\Phi(\rho,t,x),
\end{equation}
\begin{equation}
\label{1.25}
K_{10}(t,x)\Phi(\rho,t,x)= \rho\Phi(\rho,t,x).
\end{equation}
Here $\Phi(\rho,t,x)$ denotes the wave function of the particle.

For the  Casimir operators $C(\rho)$ and 
\begin{equation}
\label{1.26}
C(t,x)=\frac{l^2}{c^2}\frac{{\partial}^2}{\partial{t}^2}+\frac{l}{c}\frac{{\partial}}{\partial{t}}-l^2e^{-2ct/l}\frac{{\partial}^2}{\partial {x}^2}
\end{equation}
we have 
\begin{equation}
\label{1.27}
C(t,x){\Phi}(\rho;t,x)=C(\rho){\Phi}(\rho;t,x).
\end{equation}

To proceed further, we must construct a quantum  equation of the particle  with the property given in (\ref{1.5}a) or in (\ref{1.5}b) above. The explicit forms of the operators $K_{0d}(t,x)$, $K_{1d}(t,x)$  show that  the operators ${\Pi_0}(\rho)$ and  ${\Pi_1}(\rho)$ in (\ref{1.24}) cannot be defined as the Hamilton and the momentum operators of the particle. In the metric (\ref{1.23})  we can construct  a  momentum operator by using  two sums  $K_{1d}+\frac{mc}{\ell}K_{01}$, and $\hat\Pi_{1}(\rho)+\frac{mc}{\ell}\rho$.  We obtain the equation 
\begin{equation}
\label{1.28}
-i\hbar\frac{\partial}{\partial{x}}\Phi(\rho;t,x)=[\Pi_{1}(\rho)+\frac{mc}{\ell}{\rho}]\Phi(\rho;t,x),
\end{equation}
which defines  the operator on the right-hand side as the momentum operator of the particle in the $\rho$-representation. 
 This operator in addition to the external field contains the operator $\rho$ multipled by the mass $m$ and the parameter $\frac{c}{\ell}$.
 For the eigenfunctions $v_{\kappa}(\rho)$ of the  operator $\Pi_{1}(\rho)+\frac{mc}{\ell}\rho$ we have
\begin{equation}
\label{1.29}
(\Pi_{1}(\rho)+\frac{mc}{\ell}\rho)v_{\kappa}(\rho)={\kappa}v_{\kappa}(\rho),
\end{equation}
where ${\kappa}$ denotes the value of the  momentum of the particle in the de Sitter spacetime. A general solution of $\Phi(\rho;t,x)$  can be written in the form 
\begin{equation}
\label{1.30}
\Phi(\rho;t,x)={\int}v_{\kappa}(\rho)f_{\kappa}(t,x)d{\kappa},
\end{equation}
where $f_{\kappa}(t,x)$ are the eigenfunctions of the operators  $C(t,x)$.

For the propagator of the particle we have the expression
\begin{equation}
\label{1.31}
{\cal K}(2,1)=\int{v_{\kappa}({\rho}_2)}f_{\kappa}(t_2,x_2)v^{*}_{\kappa}({\rho}_1)f^{*}_{\kappa}(t_1,x_1)d\kappa.
\end{equation}
\section{Motion in a four dimensional dS spacetime}

The operators $H(\rho,{\bf n})$ and  ${\bf P}(\rho,{\bf n})$ in (\ref{1.5})   have the form $(spin=0)$
\begin{equation}
\label{1.32}
H(\rho,{\bf n})=mc^2\cosh\left(\frac{i\hbar}{mc}{\partial}_{\rho}\right)+\frac{i\hbar{c}}{{\rho}}{\sinh}\left(\frac{i\hbar}{mc}{\partial}_{\rho}\right) +\frac{{\bf L}^2({\bf n})}{2m{\rho}^2}e^{\frac{i\hbar}{mc}{\partial}_{\rho}},
\end{equation} 
\begin{equation}
\label{1.33}
{\bf P}={\bf n}H/c-\frac{mc}{\rho}e^{{\frac{i\hbar}{mc}{\partial}_{\rho}}}{\bf N}(\rho,{\bf n}).
\end{equation}
Here ${\bf L}({\bf n})$ and ${\bf N}(\rho,{\bf n})={\rho}{\bf n}+({\bf n}\times{\bf L}-{\bf L}\times{\bf n})/2mc$ are the operators of the Lorentz algebra in the $\rho{\bf n}$-representaion. For a particle in an external field which corresponds to a repulsive force  we use the operators 
\begin{equation}
\label{1.34}
{\Pi_0}(\rho,{\bf n})=H(\rho,{\bf n})+H'_0(\rho,{\bf n}),
\end{equation}
\begin{equation}
\label{1.35}
{\Pi_i}(\rho,{\bf n})=P_i(\rho,{\bf n})+P'_i(\rho,{\bf n}),
\end{equation}
where
\begin{equation}
\label{1.36}
H'_0=-\frac{mc^2}{2{\ell}^2}\left(\rho-i\frac{\hbar}{mc}\right)^2e^{-i\frac{\hbar}{mc}{\partial}_{\rho}},{\quad}P'_i= -n_i\frac{mc}{2{\ell}^2}\left(\rho-i\frac{\hbar}{mc}\right)^2e^{-i\frac{\hbar}{mc}{\partial}_{\rho}}.
\end{equation}
Here the quantity $\ell$ is related to the radius of a four dimensional de Sitter spacetime.

We have
\begin{equation}
\label{1.37}
{H'}^2_0-c^2{{\bf P}'}^2=0.
\end{equation}
Thus we see that the external field corresponds to  a massless field.
The operators $H'_0$,  ${\bf P}'$,  ${\bf L}({\bf n})$ and ${\bf N}(\rho,{\bf n})$ satisfy the commutations relations of the Poincar\'e algebra.

For the operators ${\Pi_0}(\rho,{\bf n})$, ${\Pi_i}(\rho,{\bf n})$ and ${\bf L}({\bf n})$, ${\bf N}(\rho,{\bf n})$ we have 
\begin{equation}
\label{1.38}
\lbrack{N_i},{\Pi_j}\rbrack=\frac{\imath\hbar}{mc^2}\delta_{ij}{\Pi_0},\quad\lbrack{\Pi_i},{\Pi_0}\rbrack=\imath\hbar{m}{\frac{c^2}{{\ell}^2}}{N_i}\quad\lbrack{\Pi_0},{N_i}\rbrack=-\frac{\imath\hbar}{m}{\Pi_i},
\end{equation}
\begin{equation}
\label{1.39}
\lbrack{\Pi_i},{\Pi_j}\rbrack=\imath\frac{\hbar}{{\ell}^2}\epsilon_{ijk}{L_k},\quad\lbrack{L_i},{\Pi_0}\rbrack=0,\quad\lbrack{\Pi_i},L_j\rbrack=\imath\hbar\epsilon_{ijk}{\Pi_k},
\end{equation}
\begin{equation}
\label{1.40}
\lbrack{L_i},{L_j}\rbrack=\imath\hbar\epsilon_{ijk}{L_k},\quad
\lbrack{N_i},{N_j}\rbrack=-\frac{\imath\hbar}{m^2c^2}\epsilon_{ijk}{L_k},\quad\lbrack{N_i},{L_j}\rbrack=\imath\hbar\epsilon_{ijk}{N_k}.
\end{equation}

The operators $\lbrace{\Pi_0},{\Pi}_i,{N_i},{L_i}\rbrace$  form a  basis for the $\rho{\bf n}$-representation of the $SO(1,4)$ group generators and correspond to constants of motion. 
The Casimir operator 
\begin{equation}
\label{1.41}
C(\rho,{\bf n})=-\frac{{\ell}^2}{(\hbar{c})^2}\left\lbrace{\Pi_0}^{2}-c^2\sum_{i=1}^{3}{\Pi}_i^{2}\right\rbrace-\frac{m^2c^2}{{\hbar}^2}{\bf N}^2+\frac{1}{{\hbar}^2}{\bf L}^2
\end{equation} 
is a multiple of the identity operator 
\begin{equation}
\label{1.42}
C(\rho,{\bf n})=(-\frac{m^2{c^2}{\ell}^2}{{\hbar}^2}-2){I}.
\end{equation}

In the four dimensional dS spacetime we use the metric which corresponds to an exponentially expanding world 
\begin{equation}
\label{1.43}
ds^2=c^2dt^2-e^{2ct/{\ell}}dx_{i}dx^{i}.
\end{equation}
Here the quantity $\ell$ is related to the radius of  the four dimensional de Sitter spacetime.

Now we introduce ten operators  which related to the Killing vectors of the spacetime (\ref{1.43}) , $(r^2=({x^i})^2)$
\begin{equation} 
\label{1.44}
K_{04}=i{\hbar}(\frac{\partial}{\partial{t}}-\frac{c{x^i}}{{\ell}}\frac{\partial}{\partial{{x^i}}}),
\end{equation}
\begin{equation} 
\label{1.45}
K_{i4}=i\frac{\hbar}{\ell}[x_i(\frac{\partial}{\partial{ct}}-\frac{{x^i}}{\ell}\frac{\partial}{\partial{{x^i}}})-\frac{\ell}{2}\frac{\partial}{\partial{x_i}}-(\frac{\ell}{2}e^{-2ct/\ell}-\frac{{r}^2}{2\ell})\frac{\partial}{\partial{x_i}}],
\end{equation}
\begin{equation} 
\label{1.46}
K_{0i}=i\frac{\hbar}{mc}[-x_i(\frac{\partial}{\partial{ct}}-\frac{{x^i}}{\ell}\frac{\partial}{\partial{{x^i}}})-\frac{\ell}{2}\frac{\partial}{\partial{x_i}}+(\frac{\ell}{2}e^{-2ct/\ell}-\frac{{r}^2}{2\ell})\frac{\partial}{\partial{x_i}}],
\end{equation}
\begin{equation}
\label{1.47}
K_{ij}=i{\hbar}(x_i\frac{\partial}{\partial{x^j}}-x_j\frac{\partial}{\partial{x^i}}).
\end{equation}
This operators satisfy the same commutation rules as  the spacetime independent  operators ${\Pi_0}(\rho,{\bf n})$, ${\Pi_i}(\rho,{\bf n})$ and ${\bf L}({\bf n})$, ${\bf N}(\rho,{\bf n})$, except for the minus signs on the right-hand sides $(d=4)$
\begin{equation}
\label{1.48}
\lbrack{K_{i0},K_{jd}}\rbrack=-\frac{\imath\hbar}{mc^2}\delta_{ij}K_{0d},\quad\lbrack{K_{id},K_{0d}}\rbrack=-\imath\hbar{m}\frac{c^2}{{\ell}^2}{K_{i0}},
\end{equation}
\begin{equation}
\label{1.49}
\lbrack{K_{0d},K_{i0}}\rbrack=\frac{\imath\hbar}{m}{K_{i4}}\quad\lbrack{K}_{id},{K}_{jd}\rbrack=-\imath\frac{\hbar}{{\ell}^2}K_{ij},\quad\lbrack{K_{ij},K_{0d}}\rbrack=0,
\end{equation}
\begin{equation}
\label{1.50}
\lbrack{K}_{id},K_{ik}\rbrack=\imath\hbar{K}_{kd},\quad\lbrack{K_{i0},K_{j0}}\rbrack=\frac{\imath\hbar}{m^2c^2}K_{ij},\quad\lbrack{K_{i0},K_{ik}}\rbrack=\imath\hbar{K_{k0}}.
\end{equation}
Just as we did in (\ref{1.24}) and (\ref{1.25}), we  write ($\Phi=\Phi(\rho,{\bf n},t,x^i)$) 
\begin{equation}
\label{1.51}
K_{0d}(t,x^i)\Phi=\Pi_0(\rho,{\bf n})\Phi,{\quad}K_{id}(t,x^i)\Phi={\Pi_i}(\rho,{\bf n})\Phi,
\end{equation}
\begin{equation}
\label{1.52}
K_{i0}(t,x^i)\Phi= N_i(\rho,{\bf n})\Phi,{\quad}K_{ij}(x^i)\Phi= {\varepsilon}_{ijk}L_k({\bf n})\Phi.
\end{equation}
These quantum equations may be used to describe a free particle in de Sitter spacetime (\ref{1.43}).

For the Casimir operators we obtain 
\begin{equation}
\label{1.53}
C(t,{x^i})\Phi=C((\rho,{\bf n})\Phi=(-\frac{m^2{c^2}{\ell}^2}{{\hbar}^2}-2)\Phi
\end{equation}
where 
\begin{equation}
\label{1.54}
C(t,{x^i})=\frac{{\ell}^2}{c^2}\frac{{\partial}^2}{\partial {t}^2}+3\frac{\ell}{c}\frac{\partial}{\partial{t}}-{\ell}^2e^{-2{\nu}t}{\nabla}^{2}.
\end{equation}

Now we have a situation fully analogous to that described in the previous section. From the equation
\begin{equation} 
\label{1.55}
i{\hbar}(\frac{\partial}{\partial{t}}-\frac{c{x^i}}{{\ell}}\frac{\partial}{\partial{{x^i}}})\Phi(\rho,{\bf n},t,{x^i})={\Pi_0}(\rho,{\bf n})\Phi(\rho,{\bf n},t,{x^i})
\end{equation}
follows that the operator ${\Pi_0}(\rho,{\bf n})$ cannot be defined as the Hamilton operator of the particle.
In this metric we can again construct the momentum operators by using   two sums \begin{equation} 
\label{1.56}
K_{i4}+\frac{mc}{\ell}K_{0i}=\Pi_{i}+\frac{mc}{\ell}N_i.
\end{equation}
We obtain  
\begin{equation} 
\label{1.57}
-i\hbar\frac{\partial}{\partial{x_i}}\Phi(\rho,{\bf n},t,{x^i})=[\Pi_{i}(\rho,{\bf n})+\frac{mc}{\ell}N_i(\rho,{\bf n})]\Phi(\rho,{\bf n},t,{x^i}).
\end{equation}
Eqs. (\ref{1.57})  defines the operators of the right-hand sides  as the momentum operators of the particle in the $\rho{\bf n}$-representation. These operators in addition to the external field contain the Lorentz boost generators $N_i(\rho,{\bf n})$ multipled by the mass $m$ and the Hubble parameter $\frac{c}{\ell}$.

 A general solution of $\Phi$ can be written in the form (${\vec\kappa}$=momentum)
\begin{equation}
\label{1.58}
\Phi(\rho,{\bf n},t,{x^i})={\int}v_{\vec\kappa}(\rho,{\bf n})f_{\vec\kappa}(t,{x^i})d{\kappa}_{1}d{\kappa}_{2}d{\kappa}_{3}
\end{equation}
where $v_{\vec\kappa}(\rho,{\bf n})$ satisfy the equation
\begin{equation}
\label{1.59}
\sum_{i=1}^{3}(\Pi_{i}+\frac{mc}{\ell}N_i)^2v_{\vec\kappa}(\rho,{\bf n})={\vec\kappa}^2v_{\vec\kappa}(\rho,{\bf n}),
\end{equation}
and $f_{\vec{\kappa}}(t,{x^i})$ are the eigenfunctions of the Casimir operator  $C(t,{x^i})$.
\section{Conclusion}
In this paper we have shown that  a generalized Schr\"odinger (GS) picture may be used to describe a free particle in a de Sitter spacetime.  In this picture the analogs of the Schr\"odinger operators of a particle are independent of both the time and space coordinates. These operators force the introduction of operators of the Killing vectors of a spacetime. The presence of this vector fields  in a quantum equation determines  the motion of a free particle in the correspondig spacetime. It was found that the spacetime independent operators with an external massless field  which corresponds to a $\it{repulsive}$ $\it{force}$ induce the operators  of Killing vectors of  the de Sitter spacetime.  The problem of determining the observables  is based  on  choosing of  the coordinate system in this  spacetime. We have shown that in the coordinates which correspond to an exponentially expanding world the spacetime independent momentum operators of the particle  in addition to the external field contain the Lorentz boost generators miltipled by the mass of the particle and the Hubble parameter. We hope that the formalism that has been presented here will be employed for solving problems in relativistic quantum physics, astrophysics and cosmology.
\section{Appendix} 
In a two dimensional Minkowski spacetime  a free relativistic particle is described by the equations (\ref{1.5}). In the  momentum representation  the Hamilton   and the momentum operator may be written in  the form ($\chi=-\ln[(p_0-cp)/mc^2]$)  
\begin{equation}
\label{1.60}
p_0={mc^2}{\cosh}{\chi},{\quad}p=mc\sinh\chi.
\end{equation}
For a particle in an external field which corresponds to a repulsive force we use the operators
 \begin{equation}
\label{1.61}
\Pi_0(\chi)=mc^2\left[\cosh\chi+\frac{1}{2}\left(\frac{\hbar}{mc{\ell}}\right)^2{e}^{\chi}\left\lbrace\frac{d^2}{d{\chi}^2}+\frac{d}{d\chi}\right\rbrace\right],
\end{equation}
\begin{equation}
\label{1.62}
\Pi_1(\chi)=mc\left[\sinh\chi+\frac{1}{2}\left(\frac{\hbar}{mc{\ell}}\right)^2{e}^{\chi}\left\lbrace\frac{d^2}{d{\chi}^2}+\frac{d}{d\chi}\right\rbrace\right],
\end{equation}
and  $N(\chi)=i\frac{\hbar}{mc}\frac{\partial}{\partial{\chi}}$.
They  satisfy the commutation relations
\begin{equation}
\label{1.63}
[N(\chi),\Pi_1(\chi)]=i\frac{\hbar}{mc^2}{\Pi_0}(\chi),\quad[{\Pi}_1(\chi),{\Pi_0}(\chi)]=i\hbar\frac{mc^2}{l^2}{N(\chi)},
\end{equation}
\begin{equation}
\label{1.64}
[\Pi_0(\chi),N(\chi)]=-i\frac{\hbar}{m}{\Pi}_1(\chi).
\end{equation}
We have 
\begin{equation}
\label{1.65}
{K}_{d0}(t,x){\Phi}(\chi;t,x)={\Pi}_{0}(\chi){\Phi}(\chi;t,x),
\end{equation}
\begin{equation}
\label{1.66}
{K}_{1d}(t,x){\Phi}(\chi;t,x)={\Pi}_1(\chi){\Phi}(\chi;t,x),{\quad}K_{01}(t,x){\Phi}(\chi;t,x)=N(\chi){\Phi}(\chi;t,x).
\end{equation}
where the function ${\Phi}(\chi;t,x)$ describe a particle in two dimensional de Sitter spacetime with metric (\ref{1.23}).

Now, from  the operators $K_{1d}+\frac{mc}{\ell}K_{01}$, and $\hat\Pi_{1}+\frac{mc}{\ell}N$ we obtain the equation 
\begin{equation} 
\label{1.67}
-i\hbar\frac{\partial}{\partial{x}}\Phi(\chi;t,x)=(\hat\Pi_{1}(\chi)+\frac{mc}{\ell}N(\chi))\Phi(\chi;t,x)
\end{equation}
which defines  the operator $\hat\Pi_{1}(\chi)+\frac{mc}{\ell}N(\chi)$ as the momentum operator of the particle  in the $\chi$-representation.
 The function $\Phi(\chi;t,x)$  can be written in the form 
\begin{equation}
\label{1.68}
\Phi(\chi;t,x)={\int}v_{\kappa}(\chi)f_{\kappa}(t,x)d{\kappa},
\end{equation}
where $v_{\kappa}(\chi)$ are the eigenfunctions of the operator  $\hat\Pi_{1}(\chi)+\frac{mc}{\ell}N(\chi)$.

\end{document}